# FORAGING BEHAVIOR OF STINGLESS BEES (*Tetragonula biroi* Friese): DISTANCE, DIRECTION AND HEIGHT OF PREFERRED FOOD SOURCE

**Authors:**
Remy R. Ciar[1]
Larry S. Bonto[1]
Mc Hale P. Bayer[1]
Jomar F. Rabajante[1,3] (corresponding author: jfrabajante@upd.edu.ph)
Sharon P. Lubag[1]
Alejandro C. Fajardo[2,3]
Cleofas R. Cervancia[2,3]

[1]Institute of Mathematical Sciences and Physics, University of the Philippines Los Baños, Philippines
[2]Institute of Biological Sciences, University of the Philippines Los Baños, Philippines
[3]UPLB Bee Program, University of the Philippines Los Baños, Philippines

## ABSTRACT

We examine the foraging behavior of stingless bees (*Tetragonula biroi* Friese) with regard to their preference of food location. The three factors considered are distance of the food source from the hive, direction of the food source, and height of the food source from the ground. Single-factor experiments for each factor were performed to determine the preferred food location. Statistical analysis shows that *T. biroi* species prefer the feeders located 1m away from the hive, which is the shortest distance used in the experiment. Results also show that this group of bees has no conclusive preferred direction, which might be a result of non-apparent cause, such as random search. In addition, the bees prefer the feeders located 1m above the ground, which are directly in front of the hive opening. Using the results from the single-factor experiments, a Three-Factor Factorial Experimental Design is used to analyze the interaction of the three factors. Factorial Analysis of Variance shows that the interaction among the three factors affects the preference of the bees. We hypothesize that the preference of the bees can be explained by (1) optimal foraging theory, (2) memory of food location, and (3) chemical marks left by conspecifics.

# INTRODUCTION

Forager bees are helpful in inducing pollination in plants. In the Philippines, bee colonies are introduced in the plantation of mangoes (*Mangifera indica* L. cv. Carabao), specifically in areas where natural pollinators are scarce. This technique increased fruit set from 3.34% to 41% [16]. Hived bee colonies are also used to pollinate coffee [13,38], cucurbits [15,22], cabbage, radish [17] and other valuable crops [5,9,27,29,37,40,61,64,67].

The famous European honeybees are not the only managed pollinators. Other types of bees, such as stingless bees, have been introduced in some Asian countries as new efficient pollinators of many commercial crops [5,16,61]. One group of stingless bee species is *Tetragonula biroi* Friese (syn *Trigona biroi*), locally known in the Philippines as *lukot* [5,6]. The use of these species as pollinators has several advantages, such as they are usually active anytime of the year, they can visit a wide variety of plants, and they can be domesticated [5,6,11].

Managing *T. biroi* is affordable. A hived colony costs about PhP2,000 - 2,500 (USD50 - 60). *Tetragonula* species need less expert management since these species do not sting. A bee colony of *T. biroi* has approximately 100,000 workers per hive which is greater than the honeybees [5,16]. In this paper, we investigate several foraging characteristics of *T. biroi* to help farmers in pollination management, especially in finding the best location of beehives so that pollination can be maximized [25].

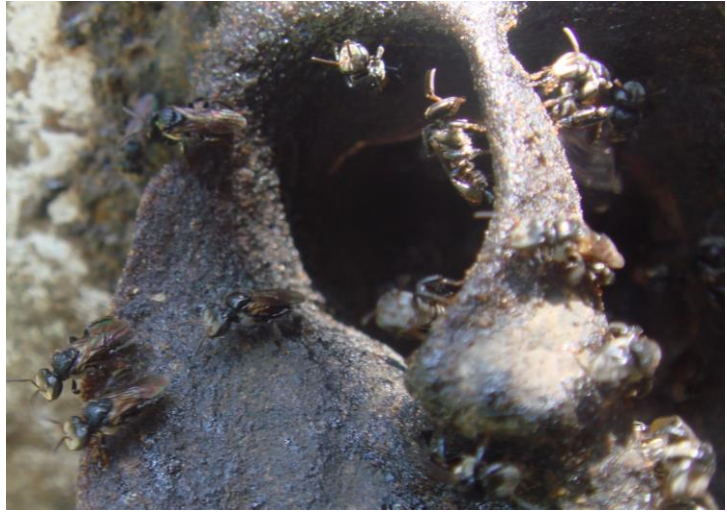

**Figure 1.** *Tetragonula biroi* Friese (Hymenoptera, Apidae, Meliponini)

In this study, we aim to determine the apparent foraging behaviors of *T. biroi* with regard to their preference of food location (using artificial feeders). We consider the distance of the food from the hive, direction of the food source (angle with respect to the compass directions; and angle with respect to the orientation of hive opening or positional directions) and height of the food source from the ground.

The maximum distance considered is 13m, although *T. biroi* can travel up to approximately 500m. This implies that this study is applicable only for small farms and greenhouse gardens. For the directionality, we only consider the cardinal (North, East, South and West) and ordinal (Northeast, Southeast, Southwest, and Northwest) directions. For the height, the maximum height is 3m. The experiments were performed only during dry weather conditions. This study was conducted from July to October 2009 within the campus and outskirts of the University of the Philippines Los Baños.

The study could help agriculturists and beekeepers in locating the suitable site to place the hive. This would contribute in maximizing crop yield and supply of food for man, and sustain biodiversity of plants.

**Bee foraging behavior, in general**

Different bee species have different criteria (via individual or collective decision-making) in choosing food sites and food quality. Climate, light intensity, wind direction and other environmental factors can affect bee foraging and floral choice [20,34,36,47,55,62,68]. Some species choose inflorescences that are near the hive [45]. There are cases where bees prefer food sources depending on the density of flowers, type of flower, and on the level of sucrose concentration [35,49,59,52,54,58]. Pollen sources also affect the foraging behavior of bees [10,63].

In some bee species, the presence of nestmates in a feeding site usually attracts foragers. Foragers deposit olfactory cues to indicate the richness (or the scarcity) of food, and these cues can attract conspecifics, not necessarily from the same colony [2,12,24,30,31,32,56,57,66].

There are bee species which use sound signals and dance language to give information about the food sources [1,7,28]. Some bees can learn patterns, such as colors, odors, shapes and structures of food [4,23,39,53,54,60].

Scouts and foragers in a social insect colony observe special recruitment and foraging systems [3,14,19,21,26,65]. Studies also found out that bees follow different theories of foraging behavior. These include Optimal Foraging Theory [48], Area

Restricted Searching Strategy [36,48], Near-Far Search [42], Levy Flights [8,50,51], and Marginal Value Theorem [48].

Obeying the Optimal Foraging Theory means that foragers are maximizing the benefits that can be obtained from the food vis-á-vis the costs needed in foraging. Benefits may include calories that can be acquired and quality of the nectar taste; while costs may include time needed in foraging, distance and height to be travelled, and competitors present near the food [48]. Marginal Value Theorem implies that foragers start to find another food source when the profitability of the food is diminishing or when the amount of food is depleting [48].

Jarau *et al.* [33] have studied foraging behaviors of stingless bees, specifically *Melipona scutellaris* and *Melipona quadrifasciata*, at food sources with different direction and distance. Nieh *et al.* [18,43,44] determined the possible mechanisms for the communication of height and distance of *Melipona panamica*. Foraging range of stingless bees can be attributed to body size [46].

# MATERIALS AND METHODS

Different experimental designs, where food locations vary in distances, directions and heights, were considered and prepared to study the foraging behaviors of *Tetragonula biroi* Friese. The study is composed of single-factor and 3-factor factorial experiments. Different bee colonies were used in all replicates.

**Training of Bees**

Bees were trained using a feeder with sucrose solution (brown sugar), an artificial food source. These bees were trained one day prior the experiment proper. Training the bees to recognize the artificial food source is important to forget previous food, to establish loyalty during experiment period, and to minimize foraging on flowers near the experiment site. The following were the procedures done in this study:

1. Preparation of the artificial food source or the feeder.
    a. 60% sucrose solutions with vanilla syrup were prepared as substitute for natural food sources. Optimal nectars among bees were 35-65% of sugar solution [52].
    b. Three tablespoons of sucrose solution were placed in each dish with cotton. Cotton was used to prevent the bees from drowning to the sucrose solution. A drop of vanilla was added to the sucrose solution to attract more bees. Each dish served as feeder for the bees.

2. Setting up the feeders.

    a. The feeders were placed at least 1m away in front of the opening of each beehive.

    b. During the food gathering of bees, the dishes were continuously being replenished with the sucrose solution.

The training location was a semi-open field with direct sunlight [47]. The training location is different from the experiment locations. Afterwards, the beehives were transferred to the experiment sites, and the entrance of each beehive was covered the night before the experiment proper. Clean dishes (without chemical marks from the bees) were used in all experiments.

## I. Single-Factor Experiments

Distance, directionality and height are the factors considered in the study. Single-factor experiments were done before analyzing the effects of interaction of the three factors. These experiments were conducted to determine the effect of each factor to the foraging activity of bees. The first single-factor experiment is concerned with distance, consisting of four replicates. This was followed by directionality and height, consisting of six and three replicates, respectively.

Trained bees were used in the single-factor experiments. Each beehive was placed on a distinct experiment site. For the distance and height experiments, the entrance of each beehive was directed towards East, since sunlight entices the bees to be active [47]. Each beehive and the corresponding food sources were placed at least 50m away from the other beehives to minimize bee visitation from the other colonies.

In every single-factor experiment, the cover of the entrance of each beehive was removed at exactly 7AM. The number of bees in each dish was observed and recorded for every 15 minutes from 7AM to 12NN. Each single-factor experiment concerning distance, directionality, or height has at least three independent replicates.

In addition, dishes were not refilled once the sucrose solution was depleted in order to detect patterns of bee transfer. Once the dishes were refilled, bees tend to stay on the same dish, which may hinder the purpose of determining the behavioral pattern of bees transferring from one feeder to another. The number of bees per feeder for every time interval was recorded using digital camera.

### I.A. Experiment Design for Distance

There are infinitely many choices for the distances that could be used in this experiment. However, this study is limited to short distances where the interval used was up to a maximum of 13m only. The shortest distance used was 1m, since beekeepers believed that beehives should not be placed too close to the food source. The distance between two posts containing the feeders was set to 2m or 3m because bees were recognized to follow "near-far" foraging search. This search suggests that bees look for food near the neighborhood of the last visited inflorescence as long as the foraging benefit is satisfactory or else foragers will go far. Two or three-meter distance between two posts is considered to be satisfactory in preventing the fuzzy impression of bees that the two posts came from the same patch of inflorescence.

Two sets of experiments were conducted. The first set was composed of feeders located at 1m, 4m, 7m, 10m and 13m away from beehive. The other set was composed of feeders located at 1m, 3m, 5m, 8m and 10m away from the beehive.

The posts containing the feeders were positioned along the direction of the hive opening forming a straight line. The height of the feeder from to the ground is the same as the height of the beehive from the ground. The height of the feeder from the ground was set constant to neglect the effect of height in the experiment. The set- up described is shown in Figure 2.

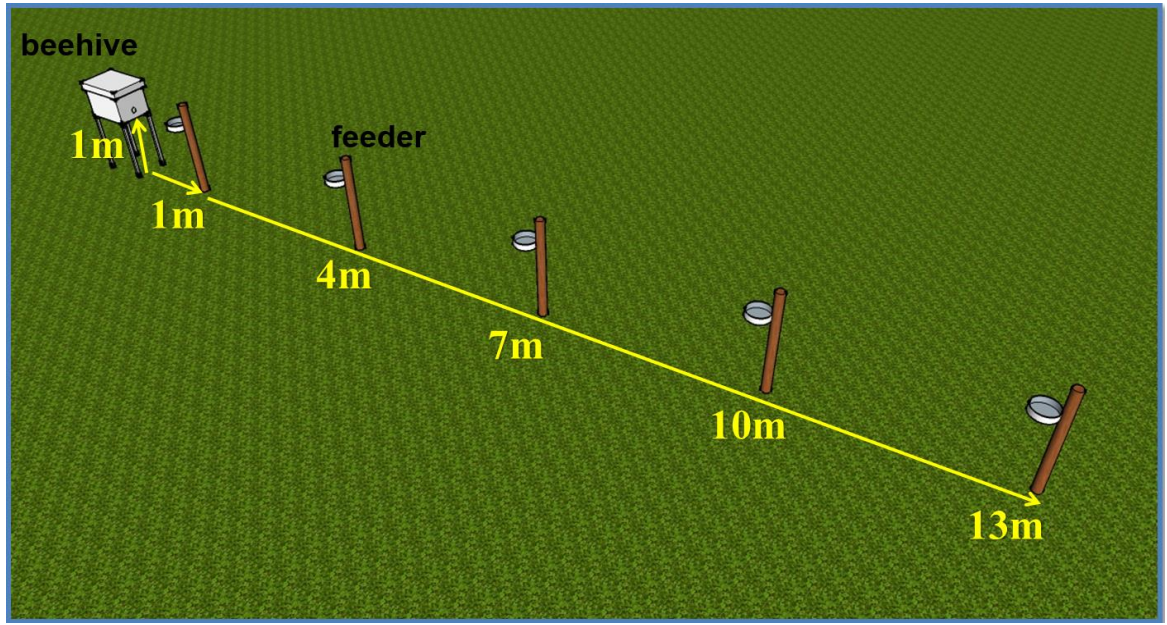

**Figure 2.** Setup of experiment for distance

This setup was placed 50m away from the training area. The location of the experiment was isolated from the training area to ensure that the bees will forage to the experiment site instead of the training site.

**I.B.    Experiment Design for Directionality**

For this experiment, eight different posts with single feeder were used. These posts were placed around the beehive representing the eight directionalities considered in the experiment. Directions consist of cardinal and ordinal directions. The distance between each post and the beehive was set to 3m.

The height of the feeder from the ground was the same as the height of the hive opening from the ground. Distance and height of feeders from the beehive were set constant to neglect their effects in the experiment. The design is illustrated in Figure 3.

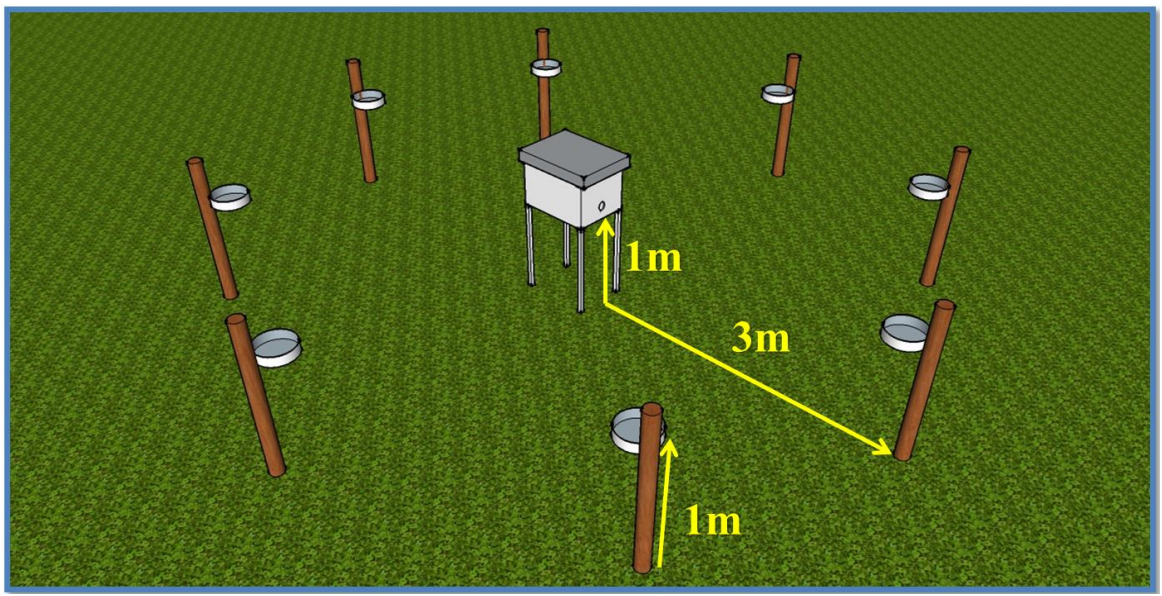

**Figure 3.** Setup of the experiment for direction

**I.C.   Experiment Design for Height**

A single post was used for this experimental design. Four feeders were placed on the post in varying heights from the ground. The heights of the feeders are 0m, 1m, 2m and 3m from the ground. The distance of the beehive from the post was set to 1m. The opening of the beehive is directly in front of the feeder located 1m above the ground. The setup is shown in Figure 4.

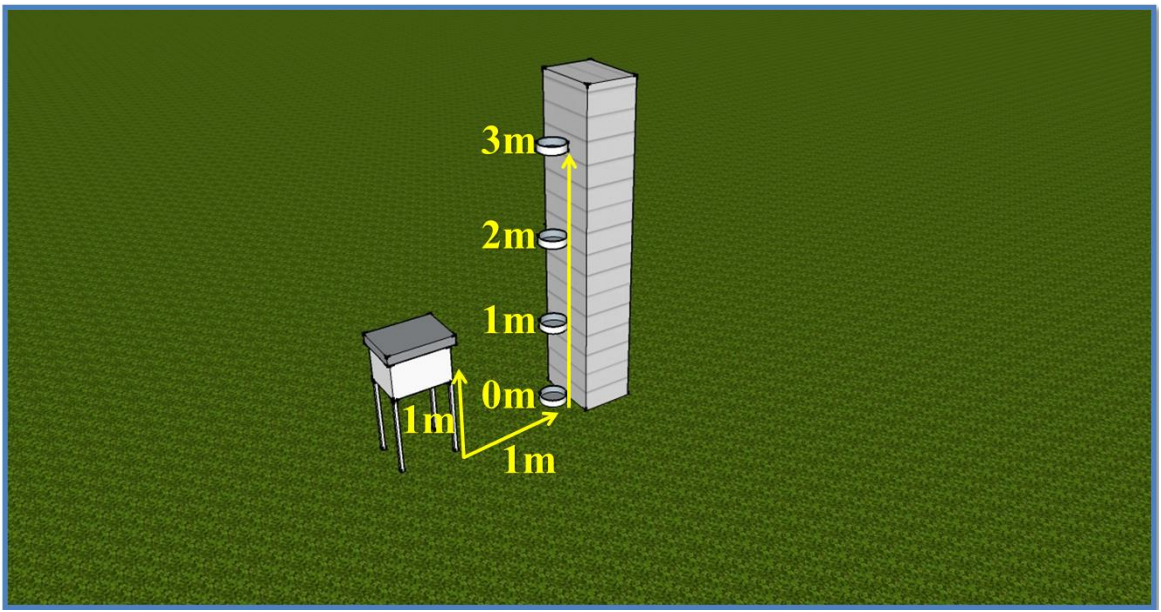

**Figure 4**. Setup of the experiment for height

**II.   Three–Factor Factorial Experimental Design**

The effects of interactions of distance, directionality, and height of feeders on the foraging behavior of *T. biroi* were examined using 3-factor factorial experiment [41]. In this factorial design, three levels for distance, four levels for directionality and three levels for height were considered. The levels considered for distance and height are the three nearest feeders to the hive considered in the single-factor experiments. Considering the three factors with different levels, the factorial design has 36 treatment combinations.

The steps for the data gathering for the 3-factor factorial experiment are the following:

1. The bees were trained one day prior the experiment proper.
2. The experiment was set up as illustrated in Figure 5.
   a. The beehive was placed in the field and its entrance was facing SE, since from the single factor experiment, this directionality is one of the preferred directionalities, and almost facing the direction of sunrise.
   b. Twelve posts were placed around the beehive. The posts were positioned in four different directions (SW, SE, NW and NE) in varying distances (1m, 4m, and 7m).
   c. Each post contains three feeders at different height levels from the ground (0m, 1m and 2m).
   d. The entrance cover of the beehive was removed at exactly 7AM.

3. The number of bees in each feeder was observed and recorded for every 15 minutes from 7AM to 12NN. The same procedure as single-factor experiment was followed in counting bees.

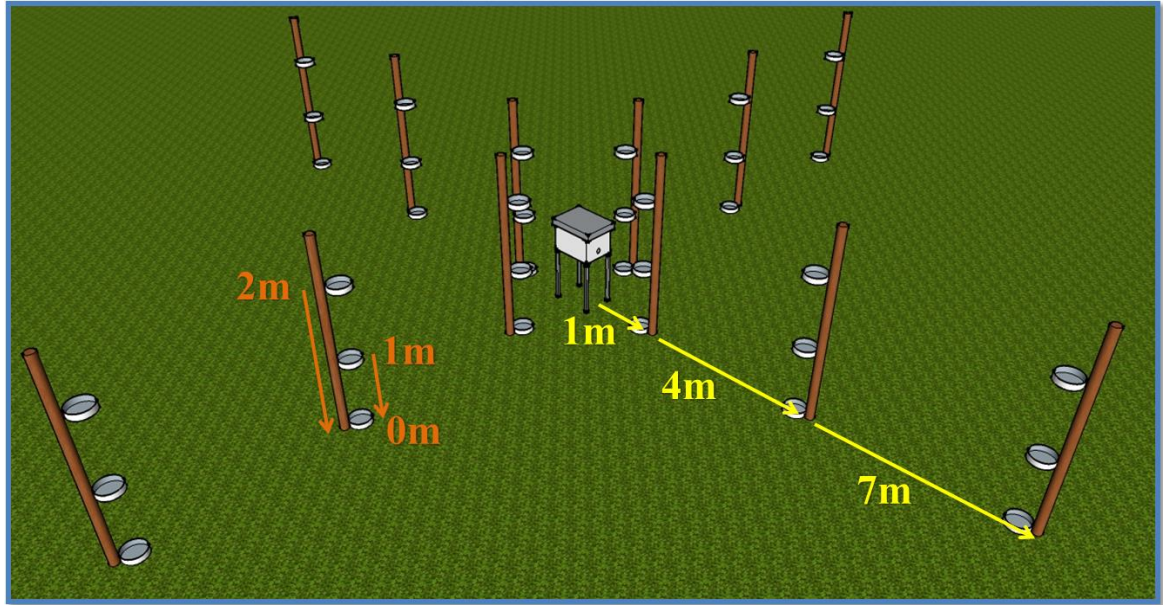

**Figure 5.** Setup of the 3-factor factorial experiment

Another two 3-factor factorial experiments with different sets of directionalities were observed. The first set consists of feeders located in N, E, W and S, with the opening of the beehive facing W, which is one of the preferred directionalities in the single-factor experiment. While, the other set consists of feeders located in of N10°E, S80°E, S10°W and N80°W, with the opening facing N10°E. N10°E is considered since there is a rule-of-thumb that beehives should be facing this angle.

# RESULTS AND DISCUSSIONS

To determine the preference of the bees with respect to food location, the data gathered from the experiments are analyzed. Two sets of frequency distributions are considered. The first set of distributions shows the number of bees that visited the feeders at varying locations. The second one is the set of distributions showing the time of visitation of bees.

The distribution of the time visitation of bees is presumed to indicate the preference of bees with respect to food location, such that, if the distribution of time visitation in Feeder A is significantly more leaning to the left compared to Feeder B, then bees prefer more the location of Feeder A. The distributions of time visitation according to the preference of bees are not normally distributed. Thus, ranks are statistically determined using the median.

It is not enough to compare the distributions of the time visitation of bees since there are cases where the relative number of bees that visited a certain feeder is negligible compared to the other feeders. To get rid of these cases, the average number of bees on each feeder is compared to the second quartile (mean) of the total number of bees that visited all the feeders during the duration of the experiment. As a rule of thumb, to be significant, the average number of bees on a feeder should be greater than or equal to the second quartile.

In the analysis of interactions, Factorial Analysis of Variance (ANOVA) – Univariate General Linear Model is used to evaluate the data resulting from the factorial experimental design. The variables are the median of the time visitation distribution. To meet the Factorial ANOVA assumptions of normality and homogeneity of variances, square-root transformation of data is used.

**I.     Single-factor Analysis**

The data gathered from each set of experiments are illustrated using bar graphs. Skewness of each graph is observed and ranked according to their position from left to right. The distribution which is more leaning to the left is assumed to be the distribution of the feeder that was first visited by bees with respect to time.

The median of the distribution of time visitation of bees would determine how early a certain feeder was visited by bees. The smaller the value of the median, the earlier is the time visitation. The feeder that was first visited by the bees would indicate the most preferred distance, directionality or height by the bees.

## I.A. Varying Distances

Looking at each graph below, it is observed that the distribution of the feeder with distance 1m is the most leaning to the left. Graphically, it could be concluded that 1m is the most preferred distance of bees. However, to give a more accurate result, two sets of distributions (number of bees and time visitation) are examined statistically.

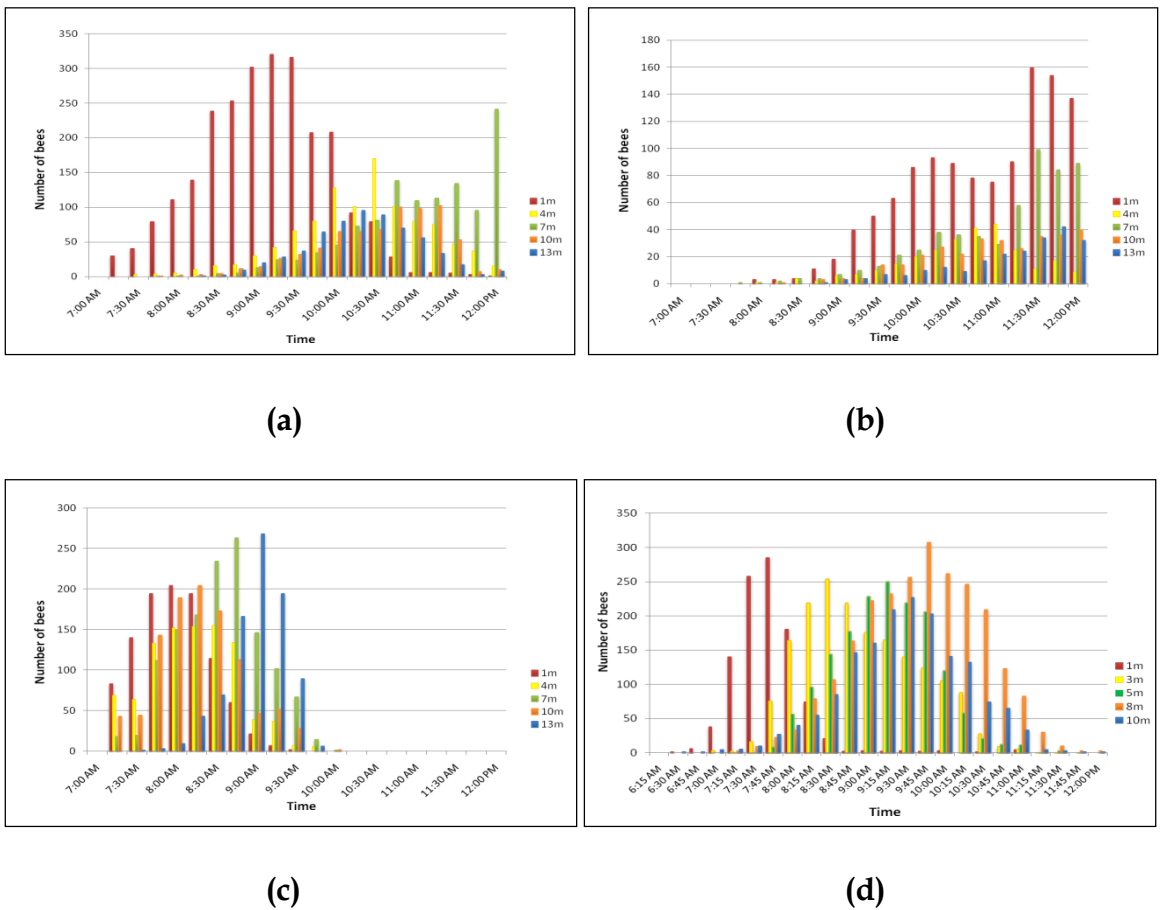

**Figure 6.** Number of bees per 15 minutes interval with respect to distance of feeder from the hive **– (a)** July 9, 2009 - colony 1, **(b)** July 9, 2009 - colony 2, **(c)** July 14, 2009 and **(d)** July 3, 2009.

### I.A.1. Distribution of Number of Bees

Using Kolmogorov-Smirnov Test of Normality, the distributions of number of visiting bees per feeder, at varying distances from the hive, are found to be not all normally distributed at 90% confidence level.

The means of the distributions of number of visiting bees are evaluated and are shown in Table 2.

**Table 2.** Means of the distributions of number of visiting bees per feeder at varying distances from the hive.

| Date of experiment | MEAN | | | | |
|---|---|---|---|---|---|
| | 1m | 4m | 7m | 10m | 13m |
| July 9, 2009 – colony 1 | 117.2857 | 48.8571 | 54.0476 | 33.4286 | 29.1429 |
| July 9, 2009 – colony 2 | 54.9524 | 12.9048 | 26.4762 | 14.8571 | 10.6190 |
| July 14, 2009 | 48.5238 | 45.2381 | 61.6190 | 49.4286 | 40.3810 |
| | 1m | 3m | 5m | 8m | 10m |
| July 3, 2009 | 50.3824 | 54.6471 | 47.7059 | 71.5000 | 49.1765 |

The mean number of bees per feeder is compared to the second quartile of the total number of bees that visited all the feeders during the duration of an experiment. This would determine whether the relative number of bees that visited a certain feeder is negligible compared to the other feeders or not. The results are shown in Figure 7.

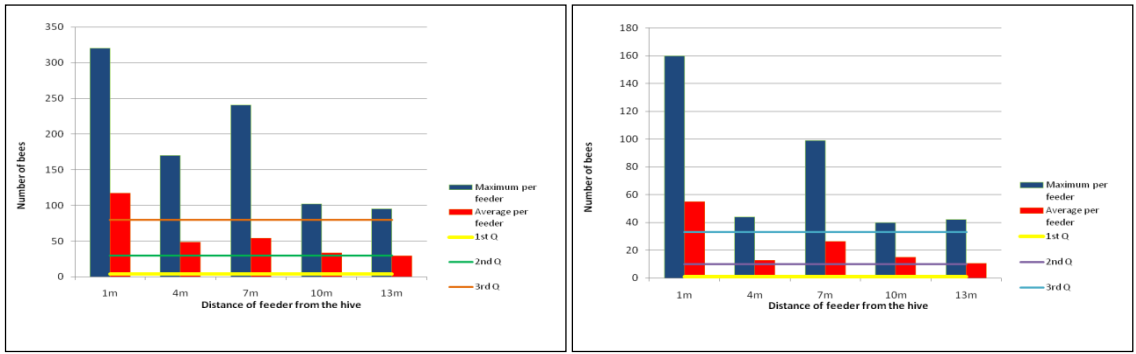

(a) (b)

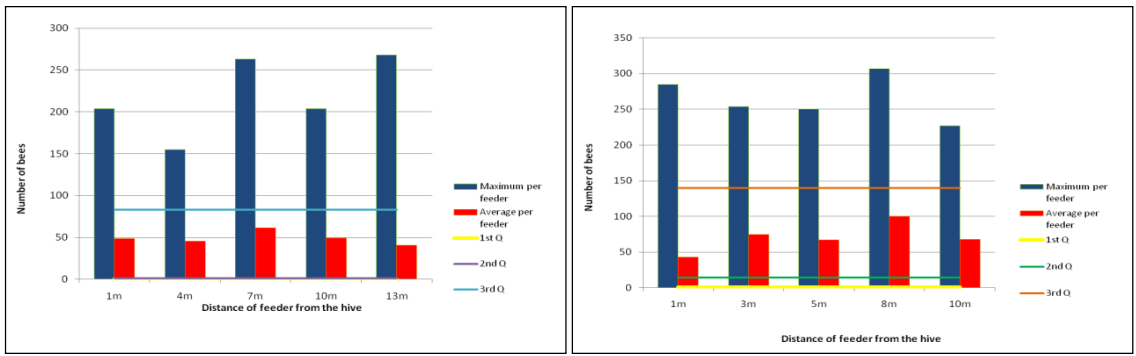

(c) (d)

**Figure 7**. Quartiles, and Maximum and Average number of bees per feeder **– (a)** July 9, 2009 - colony 1, **(b)** July 9, 2009 - colony 2, **(c)** July 14, 2009 and **(d)** July 3, 2009.

Examining every graph in Figure 7, all the mean number of bees per feeder are greater than or equal to the second quartile of the total observations except for the feeder located with distance 13m in first replicate of experiment done on July 9, 2009.

**I.A.2. Distribution of Time Visitation**

At 90% confidence level, the distributions of time visitation of bees per feeder at varying distances from the hive are found to be not all normally distributed using Kolmogorov-Smirnov Test of Normality.

Table 3 shows the medians of the distributions of time visitation of bees per feeder, which are used in comparing the distributions.

**Table 3.** Medians of the distributions of time visitation of bees per feeder at varying distances from the hive.

| Date of experiment | MEDIAN | | | | |
|---|---|---|---|---|---|
| | 1m | 4m | 7m | 10m | 13m |
| July 9, 2009 – colony 1 | 10.0000 | 15.0000 | 18.0000 | 16.0000 | 14.0000 |
| July 9, 2009 – colony 2 | 17.0000 | 16.0000 | 18.0000 | 17.0000 | 18.0000 |
| July 14, 2009 | 5.0000 | 6.0000 | 7.0000 | 6.0000 | 9.0000 |
| | 1m | 3m | 5m | 8m | 10m |
| July 3, 2009 | 7.0000 | 11.0000 | 13.0000 | 15.0000 | 14.0000 |

Since the distributions of time visitation of bees are not normal, nonparametric Wilcoxon Signed Rank Test is used to compare their medians. At 90% confidence level, the resulting comparisons for each set of distributions are the following:

a. July 9, 2009 (colony 1) – 1m < 13m < 4m < 10m < 7m

b. July 9, 2009 (colony 2) – 4m < 10m = 1m < 7m = 13m

c. July 14, 2009 – 1m < 4m = 10m < 7m < 13m

d. July 3, 2009 – 1m < 3m < 5m < 10m < 8m

Considering the results of the Wilcoxon Signed Rank Test, the distribution of time visitation of bees in the feeders located 1m away from the hive mostly have the least median time compared to the other feeders.

Therefore, it can be concluded that 1m is the most preferred distance, which happens to be the shortest distance used in the experiment. This follows the Optimal Foraging Theory, where bees forage to food locations nearer to the beehive in order for the bees to minimize energy expense.

## I.B.   Varying Directionality

The following are the beehive orientations used in different experiments:

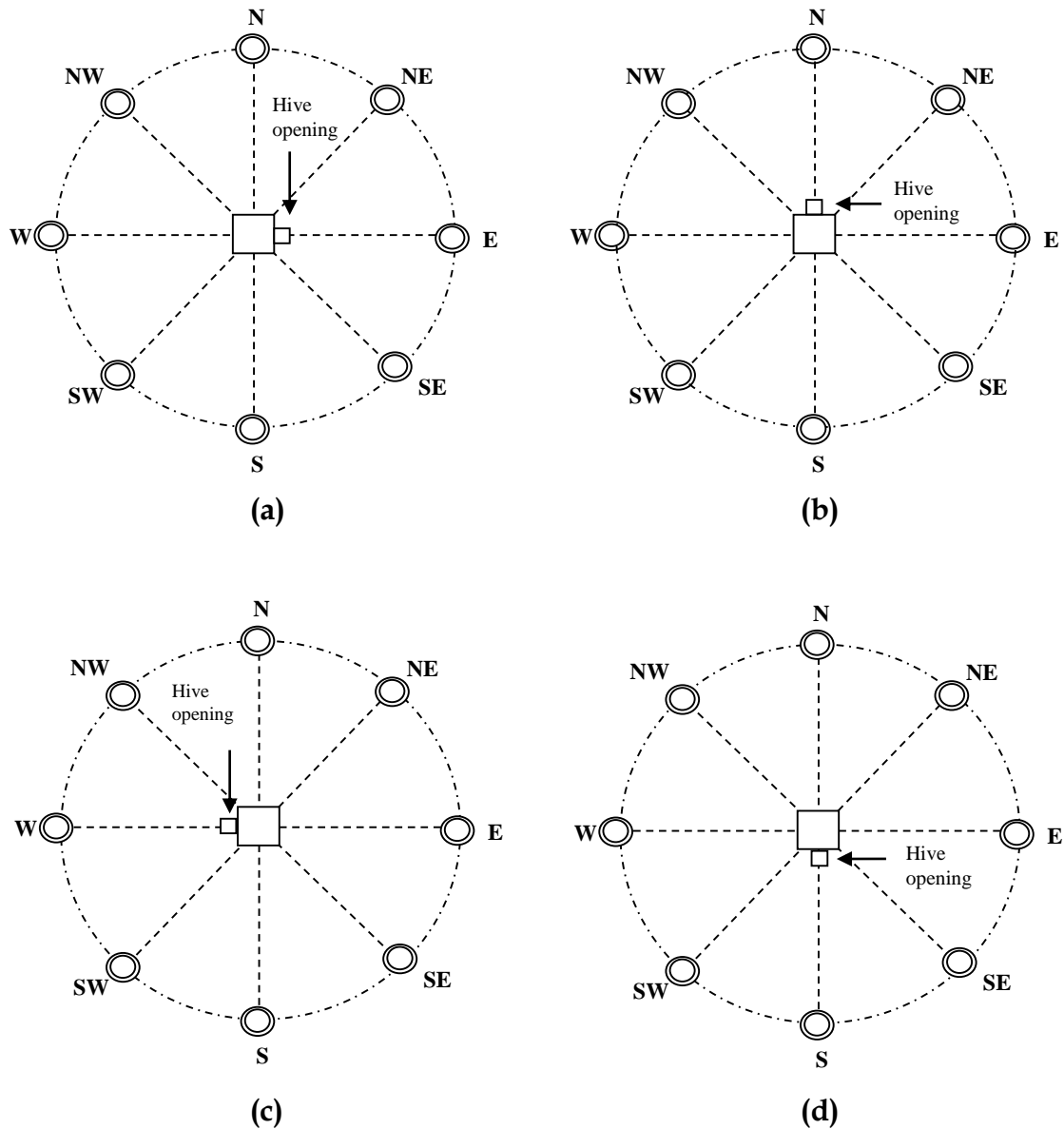

**Figure 8.** Setup for experiment – **(a)** July 10, 2009 – 3 replicates, **(b)** July 29, 2009 – colony 1, **(c)** July 29, 2009 – colony 2, and **(d)** July 29, 2009 – colony 3.

Considering the graphs of the distributions of the feeders in Figure 9, it could be observed that seemingly there are no distinct patterns. In graphs **(a), (b), (c), (d), (e)** and **(f),** the distributions of the feeders that are more leaning to the left are the feeders located at S, W, SW, SW, SE, and W, respectively. Based on the graphs, we have a hint that there is no definite directionality preferred by the bees.

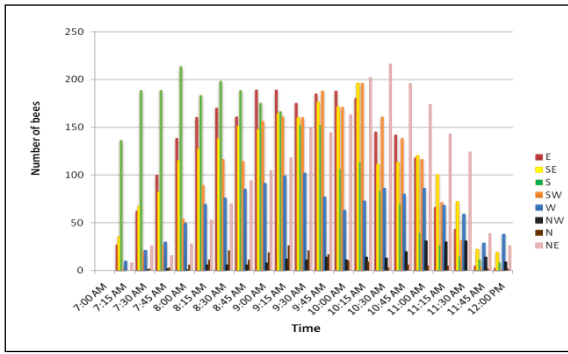

(a)

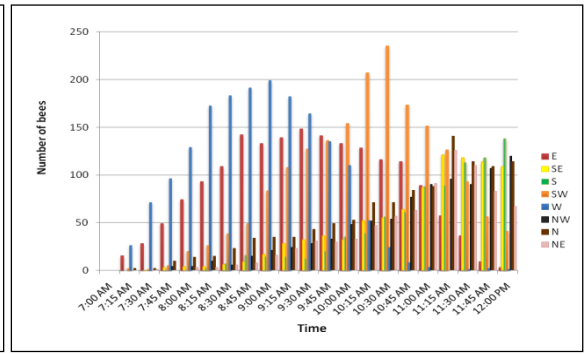

(b)

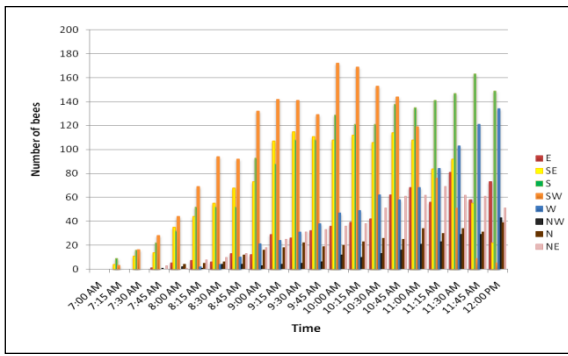

(c)

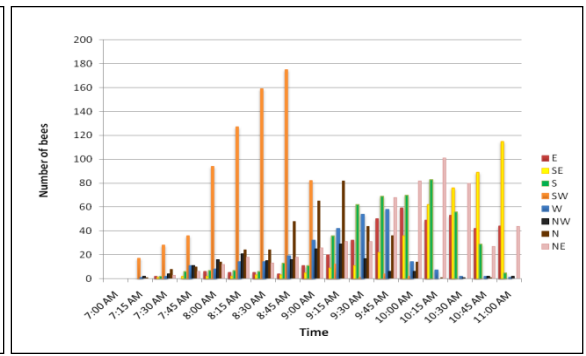

(d)

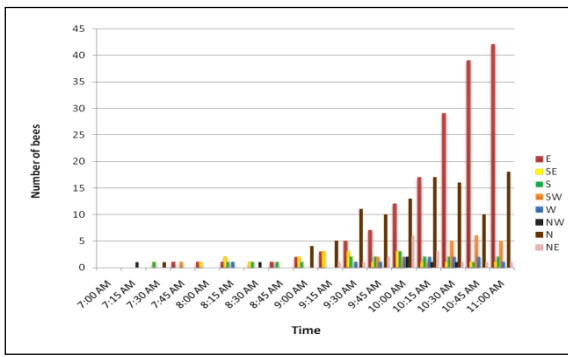

(e)

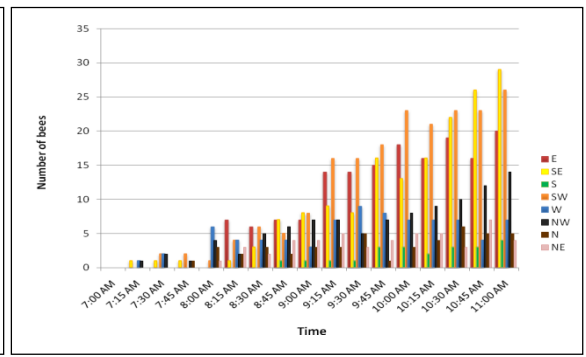

(f)

**Figure 9.** Number of bees per 15 minutes interval with respect to directionality of feeder from the hive – **(a)** July 10, 2009 – colony 1, **(b)** July 10, 2009 – colony 2, **(c)** July 10, 2009 – colony 3, **(d)** July 29, 2009 – colony 1, **(e)** July 29, 2009 – colony 2, **(f)** July 29, 2009 – colony 3.

### I.B.1. Distribution of Number of Bees

At 90% confidence level, the distributions of number of visiting bees per feeder at varying directionalities from the hive are found to be not all normally distributed using Kolmogorov-Smirnov Test of Normality.

Table 4 shows the means of the distributions of number of bees per feeder for each set of experiment.

**Table 4.** Means of the distributions of number of visiting bees per feeder at varying directionalities from the hive.

| Date of experiment | MEAN | | | | | | | |
|---|---|---|---|---|---|---|---|---|
| | E | SE | S | SW | W | NW | N | NE |
| **July 10, 2009 –** | | | | | | | | |
| colony 1 | 116.43 | 108.86 | 114.67 | 94.05 | 61.48 | 11.43 | 8.52 | 99.71 |
| colony 2 | 83.62 | 42.48 | 39.19 | 87.19 | 83.29 | 41.86 | 52.71 | 38.00 |
| colony 3 | 30.76 | 68.48 | 89.33 | 85.14 | 40.76 | 10.76 | 17.33 | 30.14 |
| **July 29, 2009 –** | | | | | | | | |
| colony 1 | 23.67 | 31.06 | 26.50 | 41.06 | 15.78 | 10.11 | 21.94 | 34.56 |
| colony 2 | 11.06 | 1.22 | 1.17 | 1.67 | 0.72 | 0.33 | 6.61 | 0.94 |
| colony 3 | 9.83 | 10.39 | 1.39 | 12.28 | 4.83 | 6.17 | 2.83 | 3.17 |

The mean number of bees per feeder is compared to the second quartile of the total number of bees that visited all the feeders during the duration of an experiment to determine whether the relative number of bees that visited a certain feeder is negligible compared to the other feeders or not. The results are shown in Figure 10.

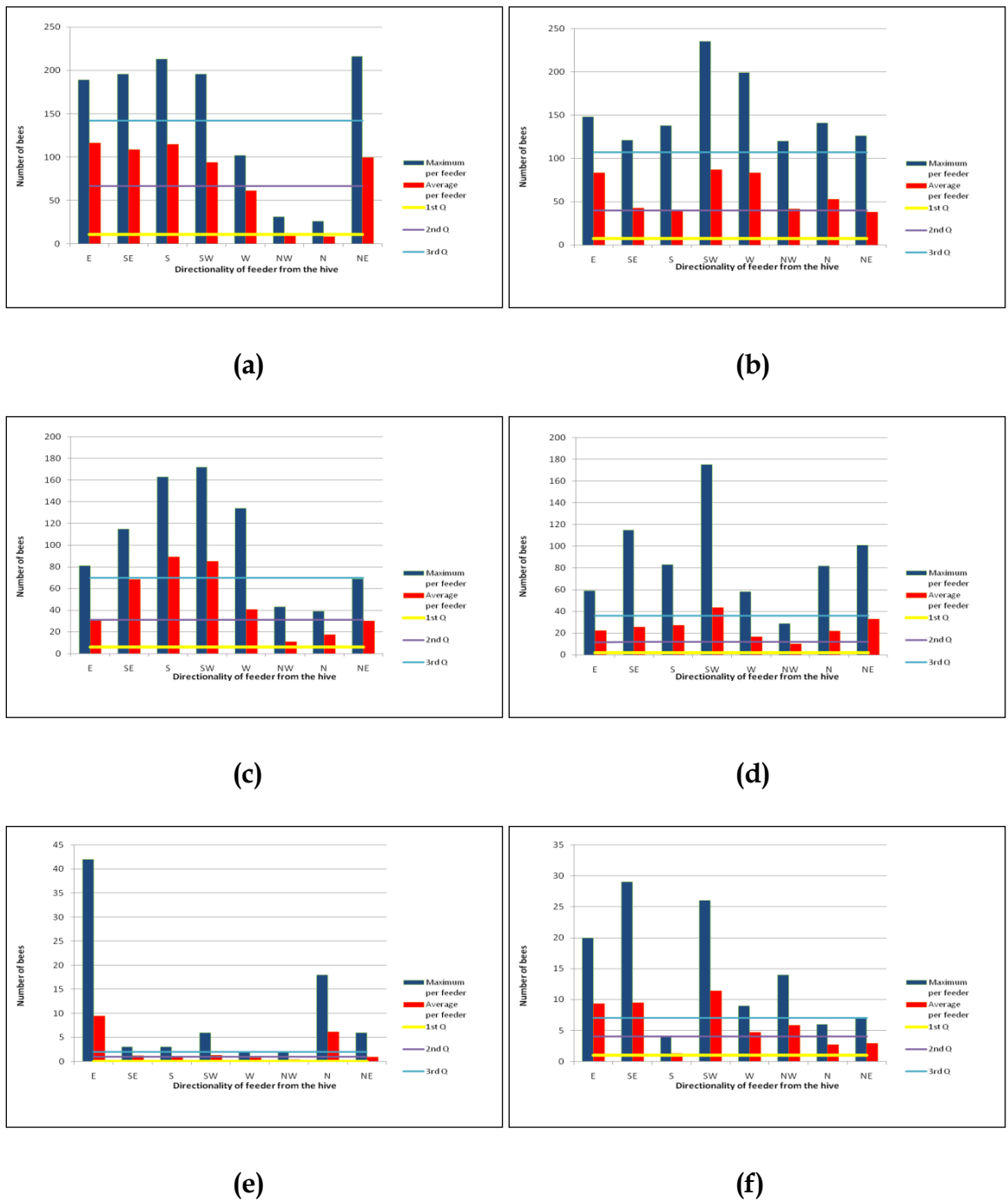

**Figure 10**. Quartiles, and Maximum and Average number of bees per feeder – **(a)** July 10, 2009 – colony 1, **(b)** July 10, 2009 – colony 2, **(c)** July 10, 2009 – colony 3, **(d)** July 29, 2009 – colony 1, **(e)** July 29, 2009 – colony 2, **(f)** July 29, 2009 – colony 3.

Examining Figure 10, in the experiment done on July 10, 2009 (colony 1) shown in graph (a), the means of the distributions of number of bees in feeders located in E, SE, S, SW and NE are all greater than or equal to the second quartile of the total observations. In the experiment done on July 10, 2009 (colony 2) shown in graph (b), feeders located in South and Northeast do not exceed the second quartile. In the experiment done on July 10, 2009 (colony 3) shown in graph (c), feeders located in Northwest, North and Northeast do not exceed the second quartile. In the experiment done on July 29, 2009 (colony 1) shown in graph (d), feeder located in Northwest does not exceed the second quartile. In the experiment done on July 29, 2009 (colony 2) shown in graph (e), feeders located in West, Northwest and Northeast do not exceed the second quartile. In the experiment done on July 29, 2009 (colony 3) shown in graph (f), feeders located in South, North and Northeast do not exceed the second quartile.

### I.B.2. Distribution of Time Visitation

At 90% confidence level, the distributions of time visitation of bees per feeder, at varying directionalities from the hive are found to be not all normally distributed using Kolmogorov-Smirnov Test of Normality.

Table 5 shows the medians of the distributions of time visitation of bees per feeder, which are used in comparing the distributions.

**Table 5.** Medians of the distributions of time visitation of bees per feeder at varying directionalities from the hive.

| Date of experiment | MEDIAN | | | | | | | |
|---|---|---|---|---|---|---|---|---|
| | E | SE | S | SW | W | NW | N | NE |
| **July 10, 2009 –** | | | | | | | | |
| colony 1 | 11.00 | 11.00 | 8.00 | 12.00 | 11.00 | 16.00 | 10.00 | 14.00 |
| colony 2 | 11.00 | 18.00 | 18.00 | 14.00 | 9.00 | 17.00 | 17.00 | 17.00 |
| colony 3 | 17.00 | 13.00 | 16.00 | 13.00 | 18.00 | 18.00 | 16.00 | 16.00 |
| **July 29, 2009 –** | | | | | | | | |
| colony 1 | 14.00 | 16.00 | 13.00 | 7.00 | 10.00 | 9.00 | 10.00 | 14.00 |
| colony 2 | 16.00 | 11.00 | 13.00 | 16.00 | 14.00 | 13.00 | 14.00 | 13.00 |
| colony 3 | 14.00 | 15.00 | 15.00 | 14.00 | 12.00 | 14.00 | 13.00 | 13.00 |

Since the distributions of time visitation of bees are not normal, nonparametric Wilcoxon Signed Rank Test is used to compare their medians. At 90% confidence level, the resulting comparisons for each set of distributions are the following:

a. July 10, 2009 (colony 1) – S < N < W = SE < E < NE < SW < NW

   The distributions of time visitation of bees in the feeder located at South, Southeast and East are the top three distributions that have the least median time.

b. July 10, 2009 (colony 2) – W < E < SW < NW < N < NE < SE < S

   The distributions of time visitation of bees in the feeder located at West, East and Southwest are the top three distributions that have the least median time.

c. July 10, 2009 (colony 3) – SW = SE < S < N = NE < E < W = NW

   The distributions of time visitation of bees in the feeder located at Southwest, Southeast and South are the top three distributions that have the least median time.

d. July 29, 2009 (colony 1) – SW < N < W = NW < S < NE = E < SE

   The distributions of time visitation of bees in the feeder located at Southwest, North and West are the top three distributions that have the least median time.

e. July 29, 2009 (colony 2) – SE < S = NE = NW < N = W < E = SW

The distributions of time visitation of bees in the feeder located at West, North and Southwest are the top three distributions that have the least median time.

The following pairwise comparisons show statistically equal median time visitation: NW=E, N=E, NW=SE, NW=S, W=SW, NW=SW, NE=SW, NW=W, NE=W, N=NW, and NE=NW.

f. July 29, 2009 (colony 3) – W < NE = N < SW = E = NW < SE = S

The distributions of time visitation of bees in the feeder located at West, Southwest, East and Northwest are the top three distributions that have the least median time.

Considering the results of Wilcoxon Signed Rank Test, there are different directionalities per experiment that can be considered as the most preferred by the bees. Therefore, no definite preferred directionality can be concluded for the single-factor experiments, indicating that the choice might be due to non-apparent behavior (such as random search) or due to environmental factors (such as significant wind blow in various directions).

Examining the locations of the feeders that are considered as the top three preferred directionalities for each set of experiment, it can be assumed that bees do not necessarily forage to feeders in front of the opening of the beehive (i.e. feeders along the direction of the opening of the hive).

## I.C. Varying Height

Looking at each graph below, it is observed that the distribution of the feeder with height 1m is the most leaning to the left. Graphically, it could be concluded that 1m is the most preferred height of bees.

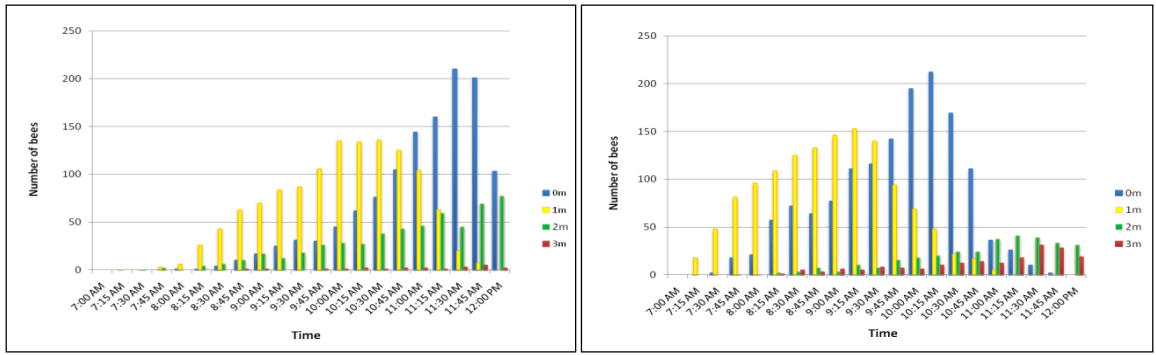

(a)          (b)

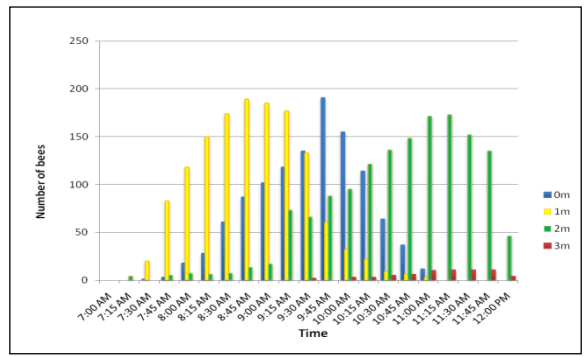

(c)

**Figure 11.** Number of bees per 15 minutes interval with respect to the height of feeder from the ground – **(a)** July 11, 2009 – colony 1**, (b)** July 11, 2009 – colony 2**,** and **(c)** July 11, 2009 – colony 3.

**I.C.1. Distribution of Number of Bees**

At 90% confidence level, the distributions of number of visiting bees per feeder at varying heights from the hive are found to be not all normally distributed using Kolmogorov-Smirnov Test of Normality.

Table 6 shows the means of the distributions of number of bees per feeder for each set of experiment.

**Table 6.** Means of the distributions of number of visiting bees per feeder at varying heights from the hive.

| Date of experiment | MEAN | | | |
| --- | --- | --- | --- | --- |
| | 0m | 1m | 2m | 3m |
| July 11, 2009 – colony 1 | 58.3333 | 57.8095 | 25.1429 | 1.0476 |
| July 11, 2009 – colony 2 | 68.6190 | 62.2381 | 15.0476 | 8.8095 |
| July 11, 2009 – colony 3 | 53.6190 | 64.8095 | 69.6667 | 3.1429 |

The mean number of bees per feeder is compared to the second quartile of the total number of bees that visited all the feeders during the duration of an experiment to determine whether the relative number of bees that visited a certain feeder is minimal compared to the other feeders. The results are shown in Figure 12**.**

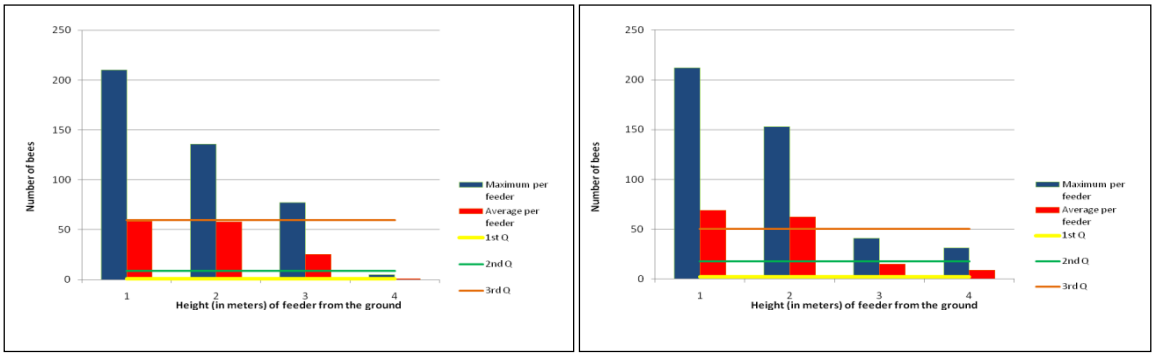

**(a)**                                                **(b)**

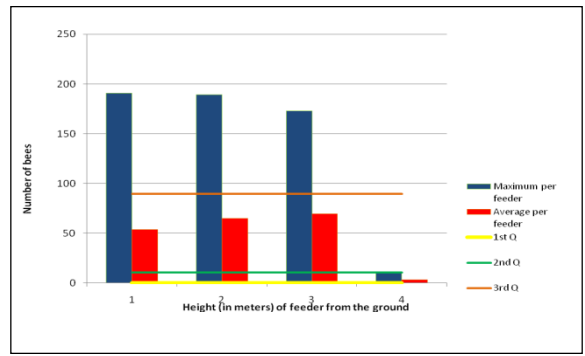

**(c)**

**Figure 12**. Quartiles, and Maximum and Average number of bees per feeder -  – **(a)** July 11, 2009 – colony 1**, (b)** July 11, 2009 – colony 2**,** and **(c)** July 11, 2009 – colony 3.

Examining Figure 12, in the experiment done on July 11, 2009 (colony 1) shown in graph (a), the means of the distributions of number of bees in feeders with height 0m, 1m and 2m are all greater than or equal to the second quartile of the total observations. This is the same with the experiment done on July 11, 2009 (colony 2) shown in graph (c). For the experiment done on July 11, 2009 (colony 3) shown in graph (b), feeders with height 0m and 1m also exceed the second quartile.

**I.C.2. Distribution of Time Visitation**

At 90% confidence level, the distributions of time visitation of bees per feeder at varying heights from the hive are found to be all not normally distributed using Kolmogorov-Smirnov Test of Normality.

Table 7 shows the medians of the distributions of time visitation of bees per feeder, which are used in comparing the distributions.

**Table 7.** Medians of the distributions of time visitation of bees per feeder at varying heights from the hive.

| Date of experiment | MEDIAN | | | |
| --- | --- | --- | --- | --- |
| | 0m | 1m | 2m | 3m |
| July 11, 2009 – colony 1 | 18.0000 | 13.0000 | 17.0000 | 17.5000 |
| July 11, 2009 – colony 2 | 13.0000 | 9.0000 | 17.0000 | 18.0000 |
| July 11, 2009 – colony 3 | 12.0000 | 8.0000 | 16.0000 | 18.0000 |

Since the distributions of time visitation of bees are not normal, nonparametric Wilcoxon Signed Rank Test is used to compare their medians. At 90% confidence level, the resulting comparisons for each set of distributions are the following:

a. July 11, 2009 (colony 1) – 1m < 2m = 3m < 0m

b. July 11, 2009 (colony 2) – 1m < 0m < 2m < 3m

c. July 11, 2009 (colony 3) – 1m < 0m < 2m < 3m

Considering each result of the Wilcoxon Signed Rank Test, the distribution of time visitation of bees at the feeder located 1m above the ground has the least median time. Therefore, it can be concluded that 1m from the ground is the most preferred height of the bees. It should be noted that this feeder has 0 degree angle of elevation from the opening of the beehive, since the feeder is placed directly in front of the opening of the beehive.

The foraging behavior of bees with regard to height also follows Optimal Foraging Theory. The feeders with 1m height from the ground have the shortest Euclidean distance from the beehive. Another hypothesis that can explain this preference is that feeders located with the same height as the hive opening can easily be discovered by scouts.

## II. Three-Factor Factorial Analysis

**Preliminary Analysis of Time Visitation of Bees per Feeder**

Different sets of legends are used in the factorial experiments (see Table 8 – 10). These legends represent the treatment combinations in the factorial experimental design. In Figure 13 – 17, line graphs are used instead of bar graphs since these graphs are visually more organized.

**Table 8.** Legend used for experiment done on August 8, 2009.

| LEGEND: | (distance, directionality, height) | LEGEND: | (distance, directionality, height) |
|---|---|---|---|
| Feeder 1 | 1m, North, 0m | Feeder 19 | 1m, South, 0m |
| Feeder 2 | 1m, North, 1m | Feeder 20 | 1m, South, 1m |
| Feeder 3 | 1m, North, 2m | Feeder 21 | 1m, South, 2m |
| Feeder 4 | 4m, North, 0m | Feeder 22 | 4m, South, 0m |
| Feeder 5 | 4m, North, 1m | Feeder 23 | 4m, South, 1m |
| Feeder 6 | 4m, North, 2m | Feeder 24 | 4m, South, 2m |
| Feeder 7 | 7m, North, 0m | Feeder 25 | 7m, South, 0m |
| Feeder 8 | 7m, North, 1m | Feeder 26 | 7m, South, 1m |
| Feeder 9 | 7m, North, 2m | Feeder 27 | 7m, South, 2m |
| Feeder 10 | 1m, East, 0m | Feeder 28 | 1m, West, 0m |
| Feeder 11 | 1m, East, 1m | Feeder 29 | 1m, West, 1m |
| Feeder 12 | 1m, East, 2m | Feeder 30 | 1m, West, 2m |
| Feeder 13 | 4m, East, 0m | Feeder 31 | 4m, West, 0m |
| Feeder 14 | 4m, East, 1m | Feeder 32 | 4m, West, 1m |
| Feeder 15 | 4m, East, 2m | Feeder 33 | 4m, West, 2m |
| Feeder 16 | 7m, East, 0m | Feeder 34 | 7m, West, 0m |
| Feeder 17 | 7m, East, 1m | Feeder 35 | 7m, West, 1m |
| Feeder 18 | 7m, East, 2m | Feeder 36 | 7m, West, 2m |

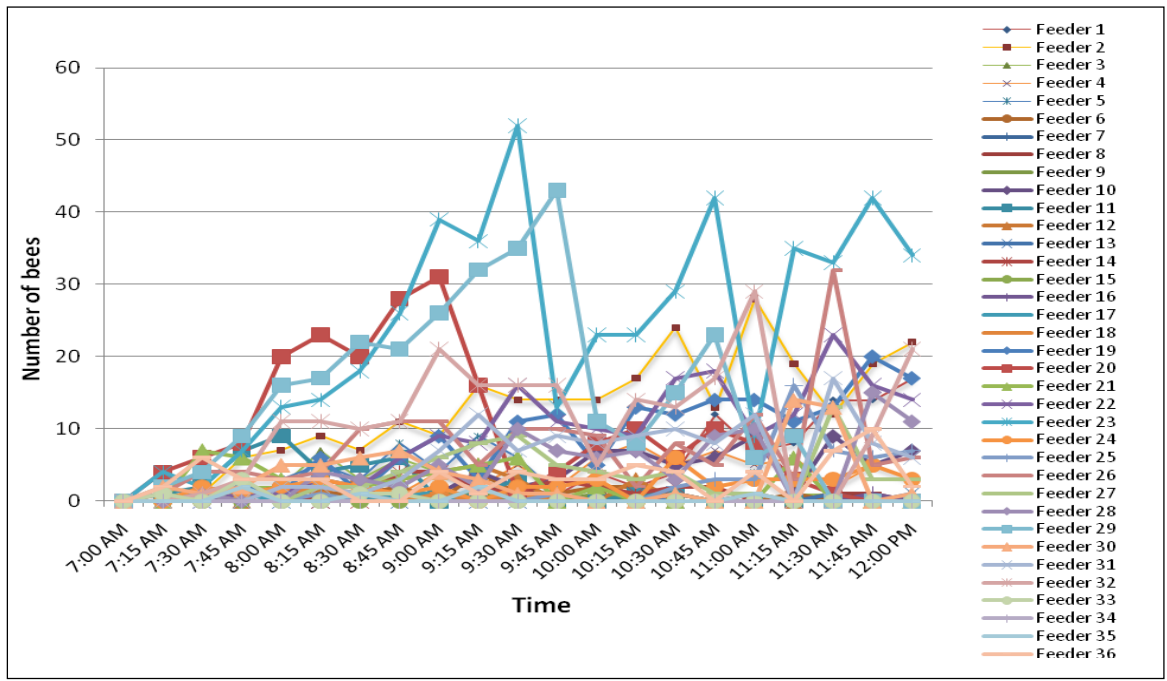

**Figure 13.** Number of bees per 15 minutes interval with respect to distance, directionality, and height of feeder (August 8, 2009).

**Table 9.** Legend used for experiment done on August 11, 12 and 14, 2009.

| LEGEND: | (distance, directionality, height) | LEGEND: | (distance, directionality, height) |
| --- | --- | --- | --- |
| Feeder 1 | 1m, Northeast, 0m | Feeder 19 | 1m, Southwest, 0m |
| Feeder 2 | 1m, Northeast, 1m | Feeder 20 | 1m, Southwest, 1m |
| Feeder 3 | 1m, Northeast, 2m | Feeder 21 | 1m, Southwest, 2m |
| Feeder 4 | 4m, Northeast, 0m | Feeder 22 | 4m, Southwest, 0m |
| Feeder 5 | 4m, Northeast, 1m | Feeder 23 | 4m, Southwest, 1m |
| Feeder 6 | 4m, Northeast, 2m | Feeder 24 | 4m, Southwest, 2m |
| Feeder 7 | 7m, Northeast, 0m | Feeder 25 | 7m, Southwest, 0m |
| Feeder 8 | 7m, Northeast, 1m | Feeder 26 | 7m, Southwest, 1m |
| Feeder 9 | 7m, Northeast, 2m | Feeder 27 | 7m, Southwest, 2m |
| Feeder 10 | 1m, Southeast, 0m | Feeder 28 | 1m, Northwest, 0m |
| Feeder 11 | 1m, Southeast, 1m | Feeder 29 | 1m, Northwest, 1m |
| Feeder 12 | 1m, Southeast, 2m | Feeder 30 | 1m, Northwest, 2m |
| Feeder 13 | 4m, Southeast, 0m | Feeder 31 | 4m, Northwest, 0m |
| Feeder 14 | 4m, Southeast, 1m | Feeder 32 | 4m, Northwest, 1m |
| Feeder 15 | 4m, Southeast, 2m | Feeder 33 | 4m, Northwest, 2m |
| Feeder 16 | 7m, Southeast, 0m | Feeder 34 | 7m, Northwest, 0m |
| Feeder 17 | 7m, Southeast, 1m | Feeder 35 | 7m, Northwest, 1m |
| Feeder 18 | 7m, Southeast, 2m | Feeder 36 | 7m, Northwest, 2m |

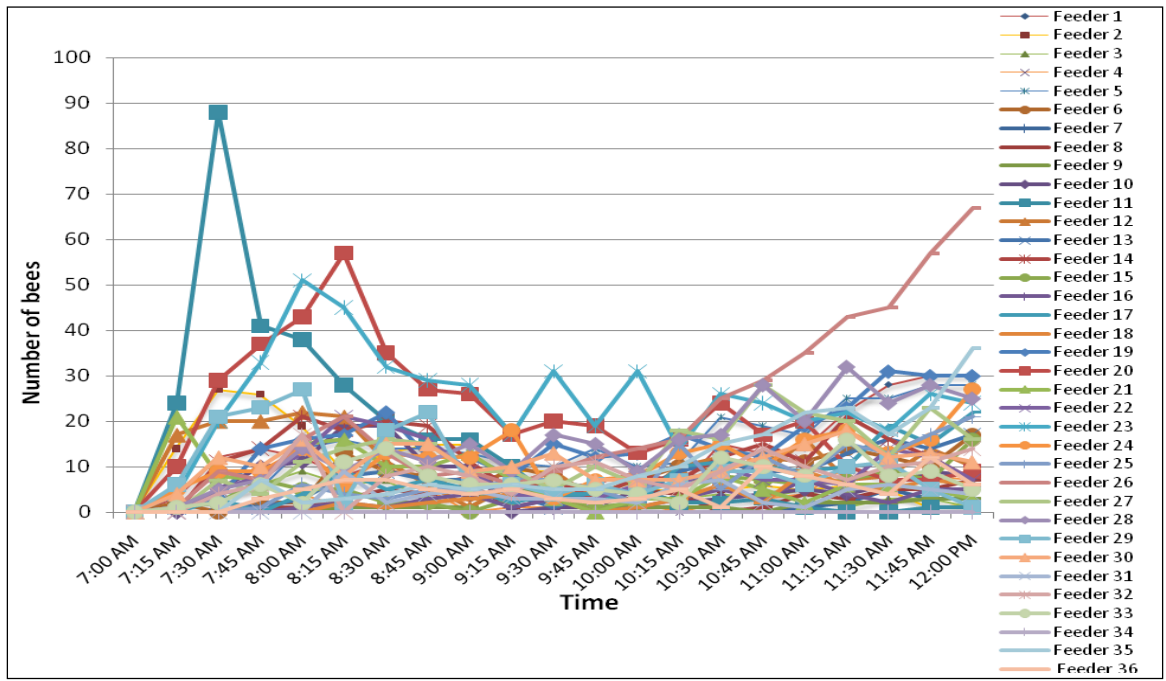

**Figure 14.** Number of bees per 15 minutes interval with respect to distance, directionality, and height of feeder (August 11, 2009).

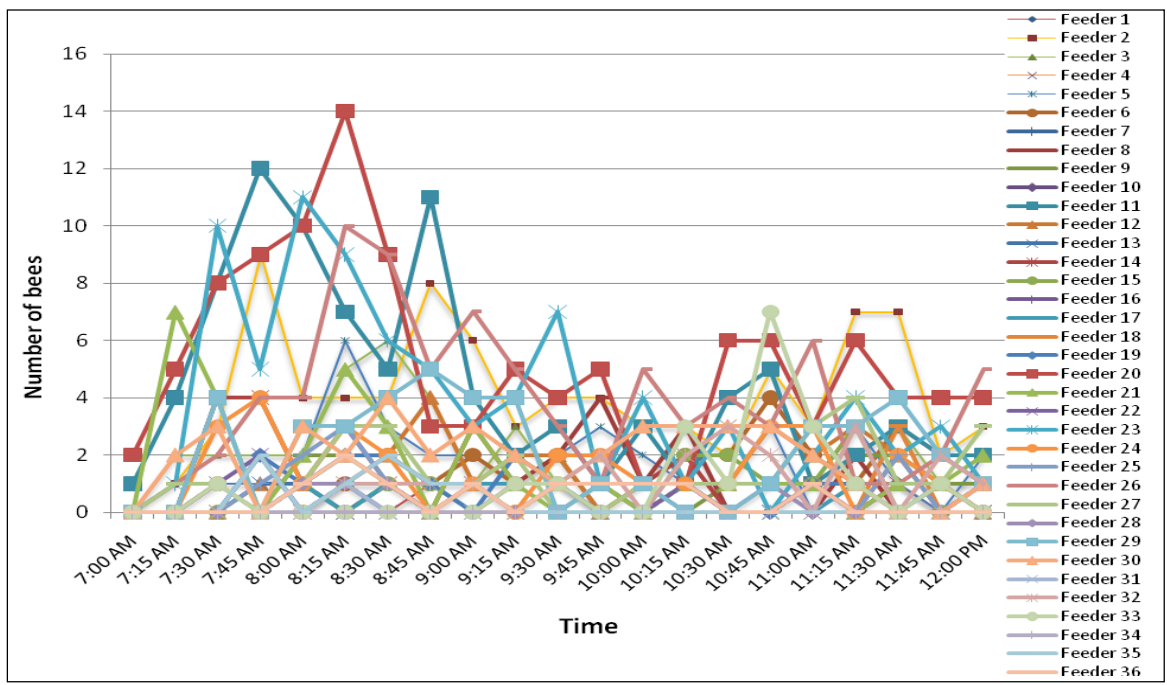

**Figure 15.** Number of bees per 15 minutes interval with respect to distance, directionality, and height of feeder (August 12, 2009).

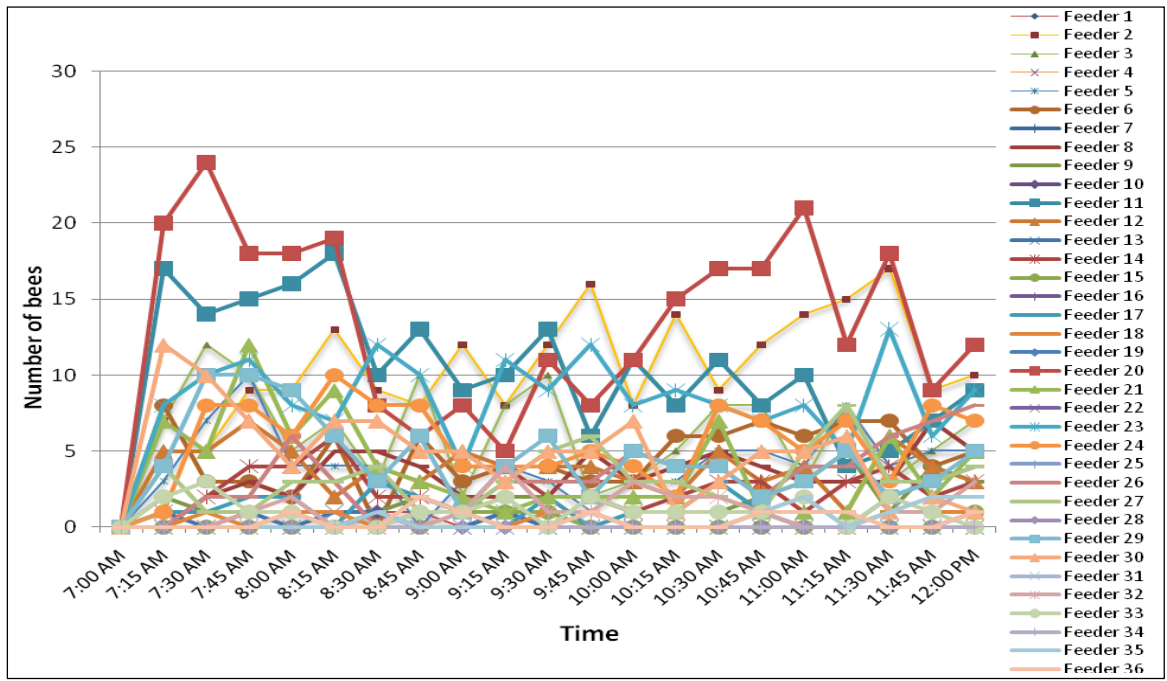

**Figure 16.** Number of bees per 15 minutes interval with respect to distance, directionality, and height of feeder (August 14, 2009).

**Table 10.** Legend used for experiment done on August 15, 2009.

| LEGEND:   | (distance, directionality, height) | LEGEND:   | (distance, directionality, height) |
|-----------|------------------------------------|-----------|------------------------------------|
| Feeder 1  | 1m, North 10° East, 0m             | Feeder 19 | 1m, South 10° West, 0m             |
| Feeder 2  | 1m, North 10° East, 1m             | Feeder 20 | 1m, South 10° West, 1m             |
| Feeder 3  | 1m, North 10° East, 2m             | Feeder 21 | 1m, South 10° West, 2m             |
| Feeder 4  | 4m, North 10° East, 0m             | Feeder 22 | 4m, South 10° West, 0m             |
| Feeder 5  | 4m, North 10° East, 1m             | Feeder 23 | 4m, South 10° West, 1m             |
| Feeder 6  | 4m, North 10° East, 2m             | Feeder 24 | 4m, South 10° West, 2m             |
| Feeder 7  | 7m, North 10° East, 0m             | Feeder 25 | 7m, South 10° West, 0m             |
| Feeder 8  | 7m, North 10° East, 1m             | Feeder 26 | 7m, South 10° West, 1m             |
| Feeder 9  | 7m, North 10° East, 2m             | Feeder 27 | 7m, South 10° West, 2m             |
| Feeder 10 | 1m, South 80° East, 0m             | Feeder 28 | 1m, North 80° West, 0m             |
| Feeder 11 | 1m, South 80° East, 1m             | Feeder 29 | 1m, North 80° West, 1m             |
| Feeder 12 | 1m, South 80° East, 2m             | Feeder 30 | 1m, North 80° West, 2m             |
| Feeder 13 | 4m, South 80° East, 0m             | Feeder 31 | 4m, North 80° West, 0m             |
| Feeder 14 | 4m, South 80° East, 1m             | Feeder 32 | 4m, North 80° West, 1m             |
| Feeder 15 | 4m, South 80° East, 2m             | Feeder 33 | 4m, North 80° West, 2m             |
| Feeder 16 | 7m, South 80° East, 0m             | Feeder 34 | 7m, North 80° West, 0m             |
| Feeder 17 | 7m, South 80° East, 1m             | Feeder 35 | 7m, North 80° West, 1m             |
| Feeder 18 | 7m, South 80° East, 2m             | Feeder 36 | 7m, North 80° West, 2m             |

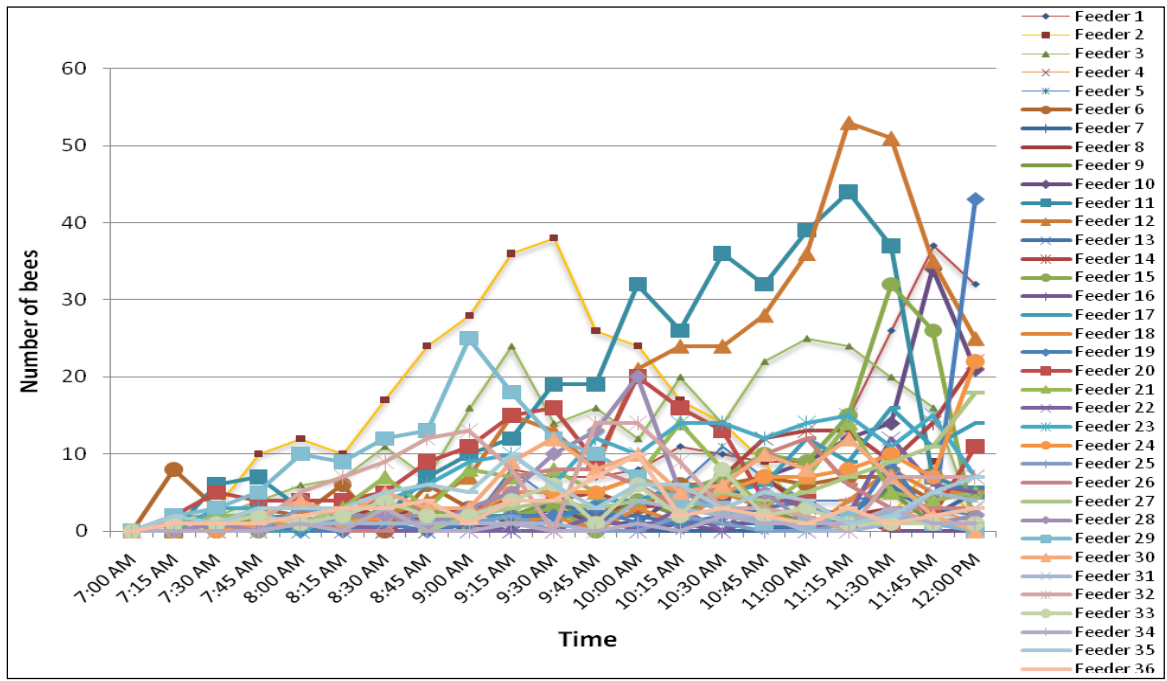

**Figure 17.** Number of bees per 15 minutes interval with respect to distance, directionality, and height of feeder (August 15, 2009).

At 90% confidence level, the distributions of number of visiting bees per feeder, for each replicate of factorial experiment, are found to be not all normally distributed using Kolmogorov-Smirnov Test of Normality. On the other hand, the mean of each feeder is compared to the second quartile of the total distribution of number of visiting bees. If the mean of a treatment combination is less than or equal to the second quartile, then the observations in this treatment are assumed to be negligible compared to the other treatment combinations.

The distributions of time visitation of bees per feeder are also tested and all are found not to be normally distributed. The medians of the distributions of time visitation of the bees are compared and the results are as follows:

a. August 8, 2009 – Feeder 7 = 15 = 34 < 33 < 11 = 18 < 31 < 17 = 20 = 21 < 35 < 5 = 12 < 9 < 6 = 14 = 30 < 16 = 29 < 27 = 32 < 8 = 26 < 36 < 13 < 2 = 4 = 22 = 23 = 24 = 31 = 28 < 10 = 19 < 1 = 25

The distribution of time visitation of bees in the feeders 11, 31, 20, 21, 5 and 12 have the least median time and these feeders are considered to be the top five preferred treatments which vary in distance, directionality and height.

The following pairwise comparisons have statistically equal median time visitation: 1=9, 1=13, 1=16, 1=17, 1=35, 2=9, 2=11, 2=13, 2=16, 2=17, 2=33, 3=5, 3=9, 3=13, 3=16, and 3=17.

b. August 11, 2009 – Feeder 34 < 11 < 2 = 21 < 10 = 12 = 20 = 29 < 3 = 22 < 23 < 14 < 30 < 8 = 9 = 32 = 33 < 19 = 24 = 31 = 36 < 5 = 6 = 27 = 28 < 1 = 13 = 15 = 16 = 17 = 25 < 18 = 26 = 35 < 4 < 7

The distribution of time visitation of bees in the feeders 11, 2, 21, 10, 12, 20 and 29 have the least median time and these feeders are considered to be the top five preferred treatments which vary in distance, directionality and height.

The following pairwise comparisons have statistically equal median time visitation: 1=22 and 3=13.

c. August 12, 2009 – Feeder 34 < 13 < 21 = 31 < 22 = 25 = 28 < 7 = 11 = 14 = 20 = 23 = 35 < 19 = 26 = 29 < 30 = 36 < 3 = 4 < 5 < 1 = 2 = 10 = 12 = 24 < 36 < 8 < 27 < 15 < 6 = 32 < 33 < 9 < 16 < 18

The distribution of time visitation of bees in the feeders 21, 11, 14, 20, 26 and 29 have the least median time and these feeders are considered to be the top five preferred treatments which vary in distance, directionality and height.

The following pairwise comparisons have statistically equal median time visitation: 1=4, 1=8, 1=9, 1=10, 1=13, 1=15, 1=17, 1=22, 1=28, 1=33, 2=4, 2=13, 2=13, 2=26 and 2=28.

d. August 14, 2009 – Feeder 16 = 19 = 22 = 28 = 34 < 13 < 7 = 4 = 21 < 1 < 29 = 30 < 10 < 11 < 3 = 12 < 15 < 14 = 23 = 24 = 33 < 18 < 2 = 5 = 20 = 36 < 27 < 8 = 17 = 32 < 6 < 26 < 9 < 31 < 35 < 25

The distribution of time visitation of bees in the feeders 21, 29, 30, 11, 3 and 12 have the least median time and these feeders are considered to be the top five preferred treatments which vary in distance, directionality and height.

The following pairwise comparisons have statistically equal median time visitation: 1=2, 1=3, 1=5, 1=6, 1=8, 1=9, 1=10, 1=11, 1=12, 1=13, 1=14, 1=15, 1=17, 1=18, 1=20, 1=21, 1=23, 1=24, 1=26, 1=27, 1=29, 1=30, 1=32, 1=33, 1=35, 1=36, 2=10, 2=15, 3=10 and 3=15.

e. August 15, 2009 – Feeder 25 < 29 < 2 < 35 < 20 = 32 = 33 = 36 < 8 = 28 = 30 < 3 = 6 = 9 = 21 < 5 = 31 < 11 = 23 = 26 < 34 < 16 < 12 = 14 = 24 < 1 = 13 = 15 = 17 = 18 = 27 < 4 = 10 = 22 < 7 < 19

The distribution of time visitation of bees in the feeders 29, 2, 35, 20, 32, 33, and 36 have the least median time and these feeders are considered to be the top five preferred treatments which vary in distance, directionality and height.

The following pairwise comparisons have statistically equal median time visitation: 1=28 and 1=36.

**Factorial Analysis of Variance**

Two sets of data are analyzed using Factorial Analysis of Variance (ANOVA). The first set consists of three replicates where the directions of feeders are according to compass directions, namely SE, SW, NW and NE. The other set consists of five replicates where the directions of feeders are according to positional directions, namely front, right, back and left of the beehive.

In the set of data with three replicates (considering compass directions), the interaction among the three factors is analyzed using Univariate General Linear Model. The result of Test of Between-Subjects Effects is as follows:

**Table 11.** Tests of Between-Subjects Effects (Compass Directions)

| Source | Type III Sum of Squares | df | Mean Square | F | Sig. |
|---|---|---|---|---|---|
| Corrected Model | 24.506 | 35 | .700 | 13.312 | .000 |
| Intercept | 1284.894 | 1 | 1284.894 | 24429.276 | .000 |
| DISTANCE * DIRECTIONALITY * HEIGHT | 24.506 | 35 | .700 | 13.312 | .000 |
| Error | 3.787 | 72 | .053 | | |
| Total | 1313.187 | 108 | | | |
| Corrected Total | 28.293 | 107 | | | |

R Squared = .866 (Adjusted R Squared = .801)

Based on the results of the Factorial ANOVA, the 3-way interaction (distance*direction*height) is significant. Since this interaction is significant, it is not necessary to place emphasis on any 2-way interactions (i.e., distance*direction, distance*height, and direction*height) and main effects that are significant. However, tests of simple (conditional) effects can be conducted to explain the interaction (e.g., using Estimated Marginal Mean Plots).

At 90% confidence level, the general linear model with the significant treatment combinations resulted from the tests of simple (conditional) effects can be represented as follows:

$$y = 3.456 - 0.314x_{112} - 0.459x_{113} - 0.436x_{121} - 0.908x_{122} - 0.354x_{123}$$
$$+ 0.458x_{131} - 0.477x_{132} - 0.876x_{133} + 0.351x_{141} - 0.571x_{142}$$
$$- 0.354x_{143} + 0.373x_{213} - 0.566x_{211} - 0.389x_{222} - 0.415x_{232}$$
$$+ 0.412x_{313} + 0.918x_{321} + 0.502x_{323} + 1.544x_{341},$$

where $x_{ijk}, i = \text{distance}(1m, 4m, 7m), j = \text{direction (NE, SE, SW, NW)}$ and $k = \text{height}(0m, 1m, 2m)$.

Considering the treatment combinations, out of twelve treatment combinations with distance 1m, only one treatment combination is not significant. Hence, distance of 1m has an effect to the median of the distributions of time visitation of bees. Moreover, most of the coefficients of the *x* variable with *i=1* in the general linear model are negative. This indicates that feeders with 1m distance from the hive are usually visited earlier.

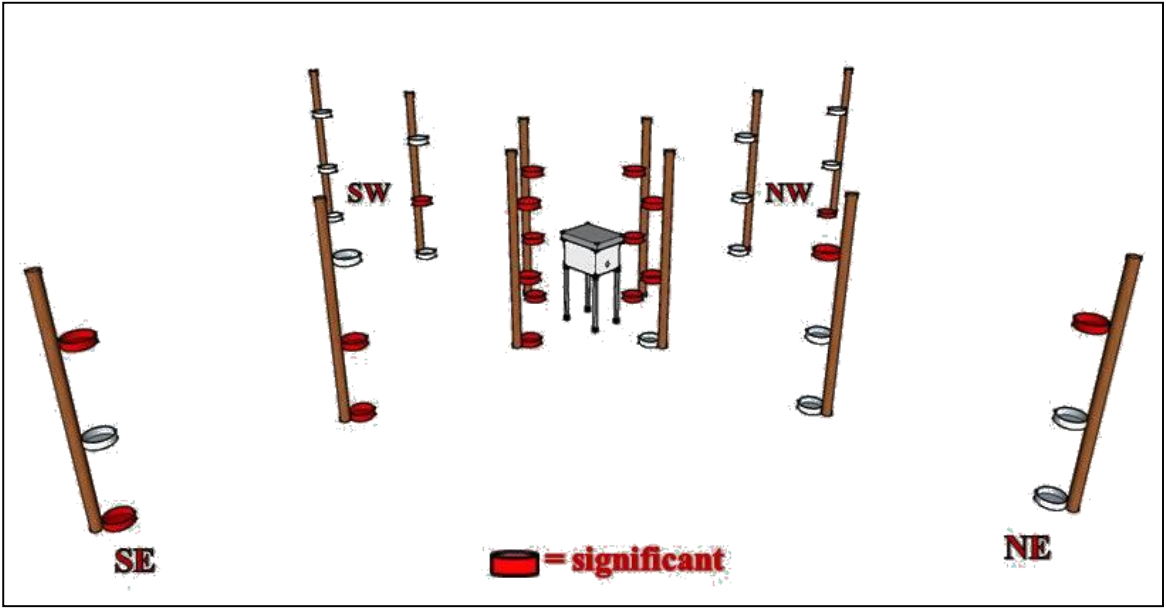

**Figure 18.** Significant Treatment Combinations (Compass Directions).

Now, in the set of data with five replicates (considering positional directions), the interaction among the three factors is analyzed using Univariate General Linear Model. The result of Test of Between-Subjects Effects is as follows:

**Table 12**. Tests of Between-Subjects Effects (Positional Directions)

| Source | Type III Sum of Squares | df | Mean Square | F | Sig. |
|---|---|---|---|---|---|
| Corrected Model | 24.718 | 35 | .706 | 11.520 | .000 |
| Intercept | 2264.738 | 1 | 2264.738 | 36942.655 | .000 |
| DISTANCE * DIRECTIONALITY * HEIGHT | 24.718 | 35 | .706 | 11.520 | .000 |
| Error | 8.828 | 144 | .061 | | |
| Total | 2298.283 | 180 | | | |
| Corrected Total | 33.546 | 179 | | | |

R Squared = .737 (Adjusted R Squared = .673)

Based on the results of the Factorial ANOVA, the 3-way interaction is significant. Since the 3-way interaction is significant, it is not necessary to place emphasis on any 2-way interaction and main effects that are significant. Tests of simple (conditional) effects can be conducted to explain the interaction (e.g., using Estimated Marginal Mean Plots).

At 90% confidence level, the general linear model with the significant treatment combinations can be represented as follows:

$$y = 3.707 - 0.271x_{111} - 0.851x_{112} - 0.464x_{113} + 0.362x_{121} - 0.370x_{122}$$
$$- 0.805x_{123} + 0.294x_{131} - 0.792x_{132} + 0.497x_{133} - 0.656x_{142}$$
$$- 0.622x_{143} - 0.326x_{211} - 0.399x_{212} - 0.349x_{213} - 0.457x_{222}$$
$$+ 0.303x_{233} + 0.812x_{311} - 0.265x_{312} + 0.320x_{321} + 0.557x_{331}$$
$$- 0.295x_{333},$$

where $x_{ijk}$, $i = \text{distance}(1m, 4m, 7m)$, $j = \text{direction (Front, Right, Back, Left)}$ and $k = \text{height}(0m, 1m, 2m)$.

Out of twelve treatment combinations with distance 1m, only one treatment combination is not significant. Thus, distance of 1m has an effect to the median of the distributions of time visitation of bees. Most of the coefficients of the *x* variable with *i=1* in the general linear model are negative. This indicates that feeders with 1m distance from the hive are usually visited earlier.

In addition, it is noticeable that seven out of nine of the *x* variable with *j=1* in the general linear model are significant and have negative coefficients. This indicates that feeders located in front of the hive opening are usually visited earlier.

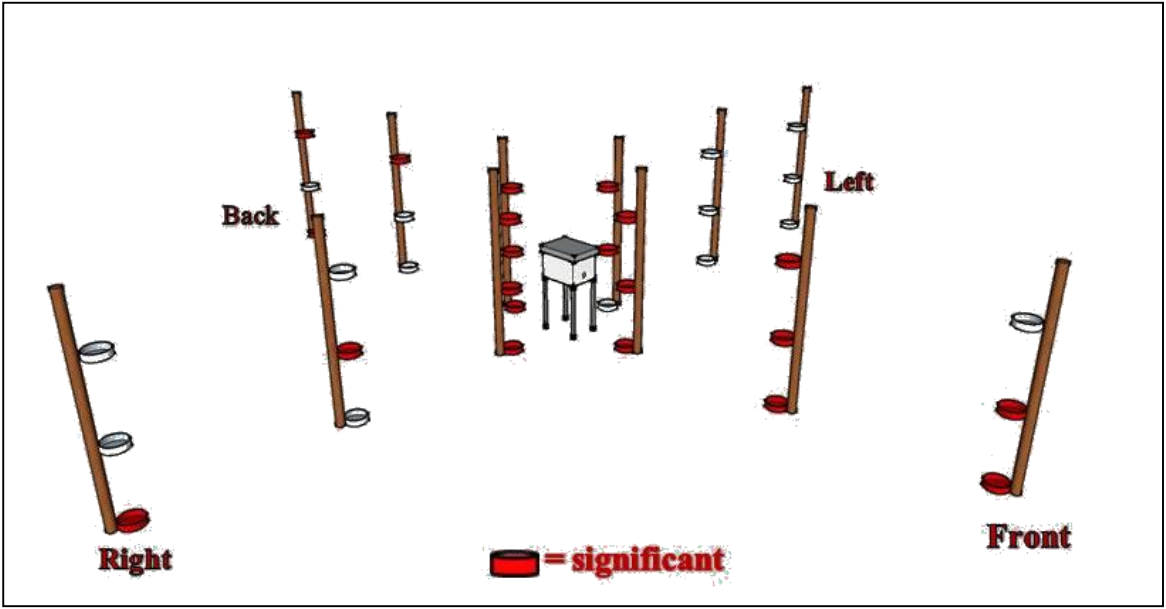

**Figure 19.** Significant Treatment Combinations (Positional Directions).

# CONCLUSION

Single-factor and three-factor factorial experiments show that *Tetragonula biroi* Friese generally follows Optimal Foraging Theory. The feeders located 1m (which is the shortest distance considered in all experiments) apart from the beehive are the most preferred by the bees. The bees forage to the nearest food sources, and usually forage to the food source in the neighborhood.

According to the results in the single-factor experiments, the food source with 1m height from the ground, having the shortest Euclidean distance from the hive opening, is the most preferred by the bees. This food source happens to be the feeder directly in front of the opening. This preference can be due to optimal foraging by bees or due to the reason that feeders located directly in front of the opening can be easily discovered by scouts.

For the directionality, results from the single-factor and 3-factor experiments show that bees do not have consistent preferred directionality. The choice of feeder varies depending on current environmental conditions (such as wind speed and direction) or might be due to non-apparent behavior (such as random search).

Area-Restricted Searching Strategy and Marginal Value Theorem can be observed in the foraging activities of *T. biroi*. Bees tend to maximize the time spent in an abundant area but when food sources are depleting they choose another feeding site which is usually in the nearest abundant neighbor. We observed that the number of bees foraging in an abundant food source continuously increases (in a nonlinear manner) but starts to decrease when the food becomes insufficient. Furthermore, we found traces of chemical marks left by foragers on the feeders. These marks indicating abundance or scarcity of food in a patch can be a communication technique used by stingless bees.

Our result regarding preference of feeders located 1m distance from and directly in front of the hive opening can also be attributed to the training done before the experiment. If this event is viable, then we can maximize the memory and learning characteristics of stingless bees. However, further investigations should be done to formulate optimal training strategies. We initially found that the training area should be a semi-open field where bees have the freedom to fly and roam around, but assures that foragers can focus on the desired food. Adequate amount of sunlight and air as well as sufficient amount of attractive food are necessary. Substantial density of foragers (simultaneously training three or more colonies in one site) can also make the training activity effective and efficient.

In determining the best location of beehive for optimizing the use of bees as pollinators, it is advisable to position the beehive at the center of the farm that will satisfy the preferred distance and height of the bees. If a beekeeper wants to pollinate a certain crop but this crop is not so attractive compared to abundant nearer plants, then the beekeeper cannot expect the crop to be maximally pollinated.

In this study, we considered short distances with maximum of 13m. The results of the study can be used for small farms and greenhouse gardens. Further investigation of bee behavior considering more combination of distance, direction and height levels during different ecological seasons can be done.

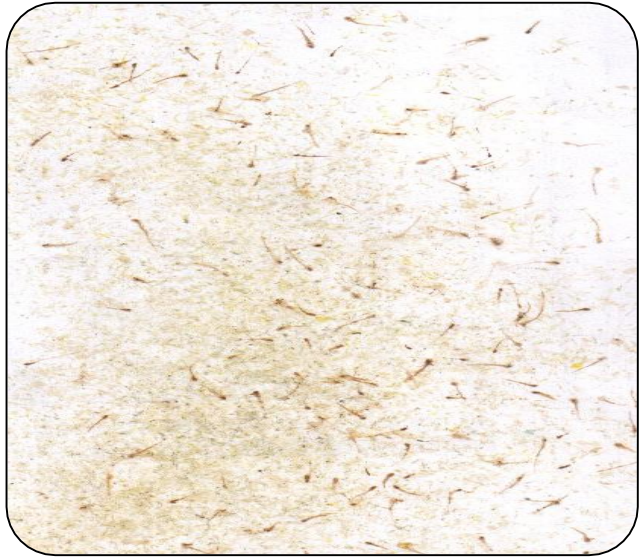

**Figure 19.** Traces of the chemical marks left by foragers.

**Acknowledgement**

We would like to thank UPLB Navigators for sharing to us their resources and place during the conduct of the experiment.

**Disclosure**

"The authors declare that there is no conflict of interests regarding the publication of this article."